\documentclass[nofootinbib,aps,preprint,longbibliography]{revtex4-1}
\pdfoutput=1

\usepackage{gensymb}
\usepackage{color}
\usepackage{graphicx}
\usepackage{epstopdf}
\usepackage{dcolumn}
\usepackage{bm}
\usepackage{amsmath}

\newcommand{\be}{\begin{equation}}
\newcommand{\ee}{\end{equation}}
\newcommand{\bfig}{\begin{figure}}
\newcommand{\efig}{\end{figure}}

\begin{document}
\title{Robust noncoplanar magnetism in band filling-tuned (Nd$_{1-x}$Ca$_x$)$_2$Mo$_2$O$_7$}
\author{Max Hirschberger$^{1,2,\dagger}$}\email{hirschberger@ap.t.u-tokyo.ac.jp}
\author{Yoshio Kaneko$^{2}$}
\author{Leonie Spitz$^{2}$}
\author{Taro Nakajima$^{2,3}$}
\author{Yasujiro Taguchi$^{2}$}
\author{Yoshinori Tokura$^{1,2,4}$}
\affiliation{$^{1}$Department of Applied Physics and Quantum-Phase Electronics Center (QPEC), The University of Tokyo, Bunkyo-ku, Tokyo 113-8656, Japan}
\affiliation{$^{2}$RIKEN Center for Emergent Matter Science (CEMS), Wako, Saitama 351-0198, Japan}
\affiliation{$^{3}$Institute for Solid State Physics, The University of Tokyo, 5-1-5 Kashiwanoha, Kashiwa, Chiba 277-8581, Japan}
\affiliation{$^{4}$Tokyo College, The University of Tokyo, Bunkyo-ku, Tokyo 113-8656, Japan}

\begin{abstract}
In the metallic pyrochlore Nd$_2$Mo$_2$O$_7$, the conducting Molybdenum sublattice adopts canted, yet nearly collinear ferromagnetic order with nonzero scalar spin chirality. The chemical potential may be controlled by replacing Nd$^{3+}$ with Ca$^{2+}$, while introducing only minimal additional disorder to the conducting states. Here, we demonstrate the stability of the canted ferromagnetic state, including the tilting angle of Molybdenum spins, in (Nd$_{1-x}$Ca$_{x}$)$_2$Mo$_2$O$_7$ (NCMO) with $x\le 0.15$ using magnetic susceptibility measurements. Mo-Mo and Mo-Nd magnetic couplings both change sign above $x=0.22$, where the canted ferromagnetic state gives way to a spin-glass metallic region. Contributions to the Curie-Weiss law from two magnetic sublattices are separated systematically.
\end{abstract}
\maketitle  

\section{Introduction}
\label{sec:intro}
Scalar spin chirality is defined in noncoplanar magnets as the triple product of nearest-neighbor magnetic moments at sites $i$, $j$, $k$
\begin{equation}
\chi_{ijk} = \mathbf{m}_i\cdot \left(\mathbf{m}_j\times\mathbf{m}_k\right)
\end{equation}
A generalization of $\chi_{ijk}$ for continuous magnetization fields $\mathbf{m}_i \rightarrow \mathbf{m}(\mathbf{r})$ is the topological winding number of skyrmions and other spin-vortex like configurations in two dimensions
\begin{equation}
\mathcal{W}(\hat{\mathbf{m}})=\frac{1}{4\pi}\int\int d^2 \mathbf{r}\, \hat{\mathbf{m}}\cdot \left(\frac{\partial\hat{\mathbf{m}}}{\partial x}\times \frac{\partial\hat{\mathbf{m}}}{\partial y}\right)
\end{equation}
Here, $\hat{\mathbf{m}}$ is a unit vector along $\mathbf{m}$ at position $\mathbf{r}$ and the area integral is over a magnetic unit cell. Previously it was shown the Hall effect of charge carriers moving through a twisted spin texture has a contribution directly proportional to $\mathcal{W}(\hat{\mathbf{m}})$, termed topological Hall effect $\sigma_{xy}^\mathrm{T}$~\cite{Bruno2004,Neubauer2009,Lee2009,Ritz2013}. More generally however, similar and closely related Hall signals arise in noncoplanar magnets whenever the lattice-weighted sum of $\chi_{ijk}$ is nonzero~\cite{Tatara2002}, as was first demonstrated experimentally using the canted, metallic pyrochlore ferromagnet Nd$_2$Mo$_2$O$_7$~\cite{Taguchi2001,Taguchi2003}. Because a quantized winding number cannot be defined in the case of Nd$_2$Mo$_2$O$_7$, the signal is referred to as geometrical Hall effect $\sigma_{xy}^\mathrm{G}$.

The established theory for $\sigma_{xy}^\mathrm{G}$ describes packets of Bloch waves moving adiabatically through a texture of local magnetic moments $\mathbf{m}(\mathbf{r})$, picking up a Berry phase as the conduction electron spin follows the the local moment texture~\cite{Bruno2004,Ritz2013}. Recently however, some of us have demonstrated that a description of $\sigma_{xy}^\mathrm{G}$ using modified band theory is more appropriate in cases where the magnetic unit cell is very small -- as is the case for the present pyrochlore magnet (Nd$_{1-x}$Ca$_x$)$_2$Mo$_2$O$_7$ (NCMO)~\cite{Hirschberger2021}. The effect of $\chi_{ijk}$ on the electronic structure was modeled in the framework of density functional theory, and it was found that (near-) touching points of bands with opposite spin character give the dominant contribution to $\sigma_{xy}^\mathrm{G}$. The new momentum-space theory was found to yield a good description of electronic transport in NCMO, where the Fermi energy $\varepsilon_F$ can be controlled by substituting Ca$^{2+}$ for Nd$^{3+}$.

It is hoped that this development represents, amongst other benefits, a step towards modeling the band structure of skyrmion hosts with nanometer-scale magnetic unit cells, as recently discovered in certain rare earth intermetallics~\cite{Kurumaji2019,Hirschberger2019,Khanh2020}.

A core assumption underlying the analysis in Ref. \cite{Hirschberger2021} is the robustness of canted magnetic order of the Molybdenum sublattice in NCMO when substituting of Ca$^{2+}$ for Nd$^{3+}$, at least for moderate values of $x\le 0.15$. Only if the canting angle $\alpha_\mathrm{Mo}\approx 5-10^\circ$~\cite{Taguchi2001,Yasui2001,Yasui2003,Yasui2006,Yasui2007} of Molybdenum magnetic moments away from their easy $\left<001\right>$ direction is fixed, systematic tuning of the Fermi energy $\varepsilon_F$ in NCMO may be compared directly to numerical calculations using the rigid band approximation. Small changes of $\alpha_\mathrm{Mo}$ cannot be easily resolved in neutron scattering~\cite{Yasui2001,Yasui2003,Yasui2006}, but it is important to confirm the stability of $\alpha_\mathrm{Mo}$ as a function of $x$, a key ingredient in the interpretation of the Hall signal.

In this article, we take a closer look at long-range order in NCMO from the viewpoint of magnetization measurements. We carefully separate two contributions to the magnetic susceptibility, and extract the exchange constants in a mean-field approximation. Using this analysis, we estimate $\alpha_\mathrm{Mo}$ in NCMO for $x\le 0.15$ (Fig. \ref{fig:alphamo}).

\section{Experimental Methods}
\label{sec:methods}


Nd$_2$O$_3$, CaCO$_3$, and MoO$_3$ powders were mixed in stoichiometric ratio and thoroughly ground to synthesize (Nd$_{1-x}$Ca$_x$)$_2$Mo$_{(4+x)/3}$O$_7$ (under evaporation of CO$_2$) by solid state synthesis in air at $1000^\circ$C over $12\,$ hours. The resulting powder was re-ground and mixed with additional elemental Mo powder to obtain the desired composition (Nd$_{1-x}$Ca$_x$)$_2$Mo$_2$O$_7$, before being heated to $800^\circ$C for half a day in argon gas. After grinding for a third time, the reaction product was pressed to form a rod and sintered in two steps ($1000^\circ$C for $12$ hours, then $1400^\circ$C for 12 hours) in argon gas. The rods were zone melted in a bespoke~\cite{Kaneko2020,Hirschberger2021} laser-floating zone furnace (LFZ) under argon flow at $1630-1700^\circ$C and at $1.8-2\,$mm/h vertical speed. Single crystals of cm-size were obtained for $x = 0.01$, $0.03$, $0.05$, $0.07$, $0.10$, $0.15$, $0.22$, $0.30$, and $0.40$.

\begin{figure}[htb]
  \centering
  \includegraphics[clip, trim=0.3cm 0.44cm 0.82cm 1.15cm, width=0.6\linewidth]{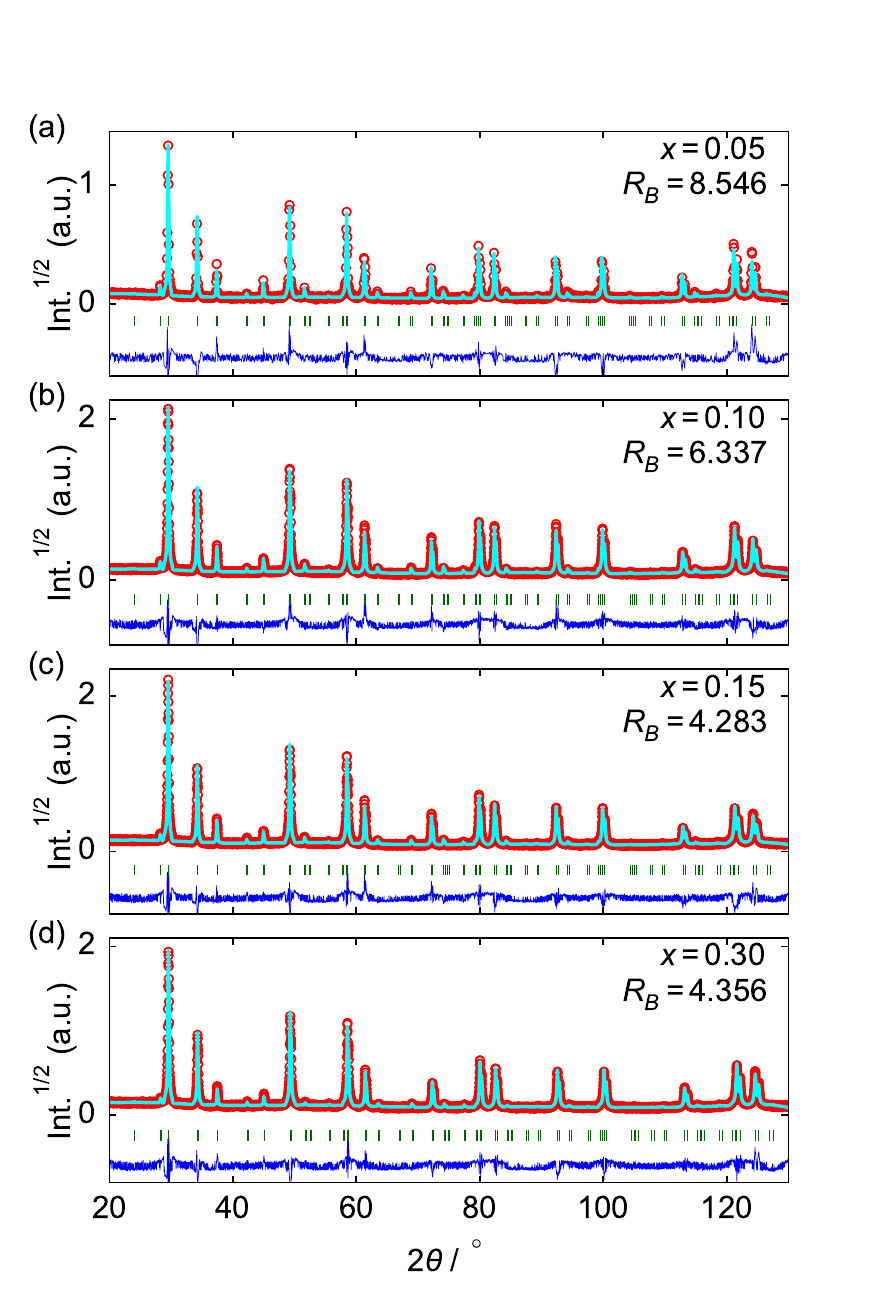}
  \caption{Rietveld refinement of powder x-ray diffraction data from crushed single-crystal pieces of (Nd$_{1-x}$Ca$_x$)$_2$Mo$_2$O$_7$ obtained at room temperature. Red dots are raw data, cyan is the fit curve, blue is the difference between the former two (shifted by an offset). The green lines indicate expected reflections from the cubic pyrochlore structure. The square root of the scattered intensity is shown. No phase impurities beyond $1\,\%$ volume fraction were detected in this analysis. $R_B$ is a figure of merit defined in the text.}
\label{fig:xrd}
\end{figure}

\begin{figure}[htb]
  \centering
  \includegraphics[clip, trim=0.0cm 0.0cm 0.0cm 0.0cm, width=0.8\linewidth]{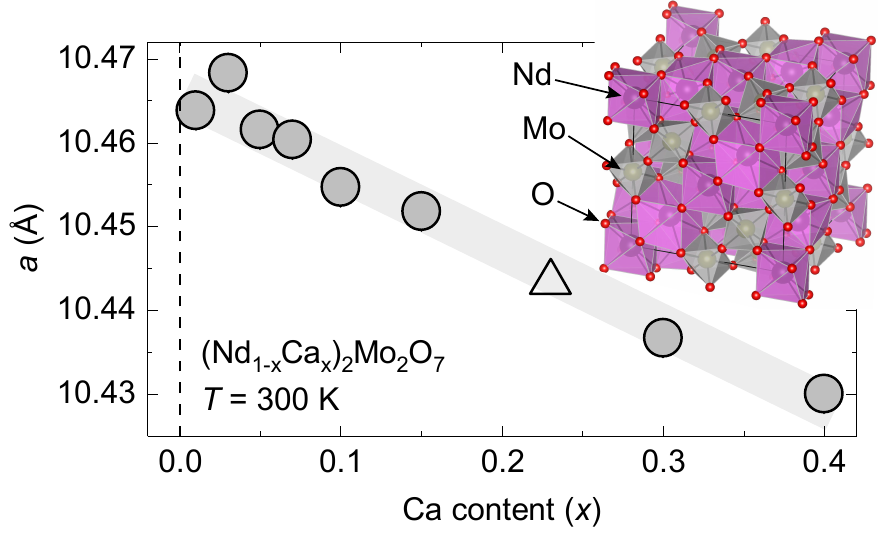}
  \caption{Cubic lattice constant $a$ for the pyrochlore structure of (Nd$_{1-x}$Ca$_x$)$_2$Mo$_2$O$_7$ and its evolution with $x$. A triangle marks a crystal batch for which the nominal Ca content differed from the measured value (see text). The grey shaded line is a guide to the eye. Statistical error bars are smaller than the symbol size. Inset: cubic pyrochlore structure of Nd$_2$Mo$_2$O$_7$ where pink and grey polyhedra mark local environments of Nd and Mo, respectively~\cite{Momma2011}. Oxygen atoms are shown in red.}
\label{fig:vegard}
\end{figure}

We characterized the resulting single crystals by examination of polished surfaces under an optical microscope with Nicol prism, verified the single-crystalline nature of individual, cm-sized pieces by Laue x-ray diffraction, and confirmed the single-phase nature by powder x-ray diffraction (XRD) of crushed single-crystalline pieces. Figure \ref{fig:xrd} shows powder patterns at selected $x$, confirming the absence of impurity phases with volume fraction above $1\,\%$. A full Rietveld refinement in the Fd$\bar{3}$m pyrochlore structure was carried out using the software suite RIETAN-FP~\cite{Izumi2007}. We show the square-root of intensity to emphasize (small) discrepancies of data and refinement, where $R_B$ is the reliability index calculated from integrated intensities of the Bragg reflections. From the fits, we obtained the cubic lattice constant $a(x)$, whose linear evolution with $x$ confirms substitutional doping of Ca$^{2+}$ on the Nd$^{3+}$ site (Vegard's law). One crystal, marked by a triangle in Fig. \ref{fig:vegard}, was grown at low temperature in self-flux condition ($T=1500^\circ$C). For this sample, nominal Ca-content $x_\mathrm{nom.}= 0.40$ and measured $x=0.22$ differed, as Ca is pushed out of the melting zone under these growth conditions. 

For in-house magnetization measurements in a Quantum Design MPMS-3 cryostat, single crystals were hand-polished into cubes of lateral dimensions $\sim 0.8\,\mathrm{mm}^3$ and mass $m\approx 2-3\,$mg. Magnetization isotherms $M(H)$ and $M(T)$ curves at fixed magnetic field $H$ were obtained using the extraction technique. For the discussion of the Curie-Weiss law at higher temperature, a moderately high $\mu_0H = 1\,$T was applied to the crystals in order to obtain reasonable signal strength despite the small size of the cube-like pieces. 

The demagnetization effect, i.e. the difference between $H$ and the internal magnetic field $H_\mathrm{int} = H-NM$, was found to be negligible for high and moderate $H$. $N$ and $M$ are the demagnetization factor calculated in elliptical approximation, and the bulk magnetization density, respectively. Note however that it was found preferable to prepare cube-shaped samples for low-field experiments in $\mu_0H = 0.01\,$T, to enhance the reproducibility of these experiments (section \ref{sec:mt}).

\section{Experimental Results}
\label{sec:experimental}

\subsection{Low-field magnetization and phase diagram}
\label{sec:mt}
At low magnetic fields, the magnetization $M(T)$ is sensitive to the onset of long-range order (Fig. \ref{fig:mt}, dashed, vertical black lines). The canted ferromagnetic (C-FM) state ($x\le 0.20$) manifests itself as a step-like increase at $T_C$, followed by a relatively flat slope of $M(T)$ at $T_C>T>20\,$K. In contrast, crystals with $x>0.20$ show a cusp-anomaly reminiscent of the behavior expected for antiferromagnets. For example at $x = 0.40$, the small value of $M$ indicates the absence of a spontaneously ordered moment (note different $y$-axis scales in Fig. \ref{fig:mt}). We tentatively identify the kink temperature $T_G$ with the onset of a spin-glass metal state (SGM) where antiferromagnetic Mo-Mo spin correlations dominate (see Discussion). 

\begin{figure}[htb]
  \centering
  \includegraphics[width=0.8\linewidth]{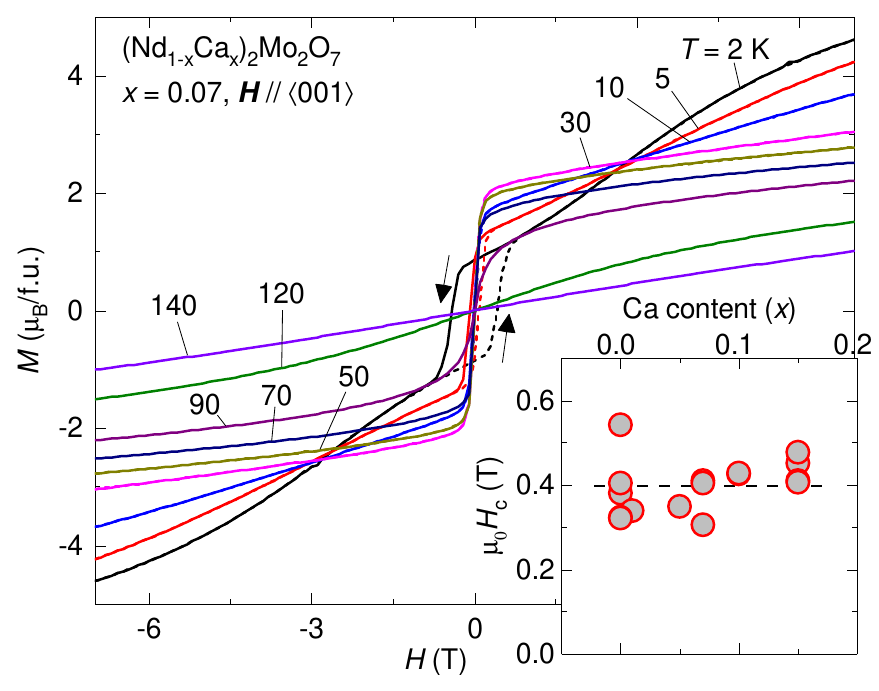}
  \caption{Magnetization isotherms $M(H)$ for (Nd$_{0.93}$Ca$_{0.07}$)$_2$Mo$_2$O$_7$. Hysteresis appears at the lowest $T<20\,$K, where dashed and solid lines mark increasing and decreasing magnetic field, respectively (c.f. black arrows). Inset: coercive field $H_c$ for external magnetic field $\mathbf{H}$ along the $\left<001\right>$ axis, at $T=2\,$K.}
\label{fig:mh}
\end{figure}

Two modes of field history gave starkly discrepant $M(T)$ signals at low temperature: In field-cooled (FC) mode, the samples are heated to $T = 150\,$K, cooled to $2\,$K in $\mu_0 H = 0.01\,$T, and the magnetization is recorded while $T$ is increased in steps. Contrast this with high-field cooled (HFC) sample preparation, where $\mu_0H = 1\,$T is applied at $150\,$K and during the cooldown, upon which the field is decreased to $0.01\,$T and again data is collected while increasing $T$ in steps. All samples have a strong divergence of FC and HFC curves below $T^*\sim 20\,$K, behavior that is associated with the rapid evolution of long-range magnetic order on the rare earth sublattice and concomitant hysteresis of the $M(H)$ curves (see Fig. \ref{fig:mh}). In an inset of the latter figure, we also show the evolution of the coercive field $H_c$ as a function of Ca content, within the boundaries of the C-FM phase. The data suggest that magnetic anisotropy of individual Nd$^{3+}$ is hardly affected by chemical dilution with Ca$^{2+}$ and consequent distortion of their local environment.

\begin{figure}[b]
  \centering
	\includegraphics[clip, trim=0.25cm 0.57cm 0.35cm 0.2cm, width=0.7\linewidth]{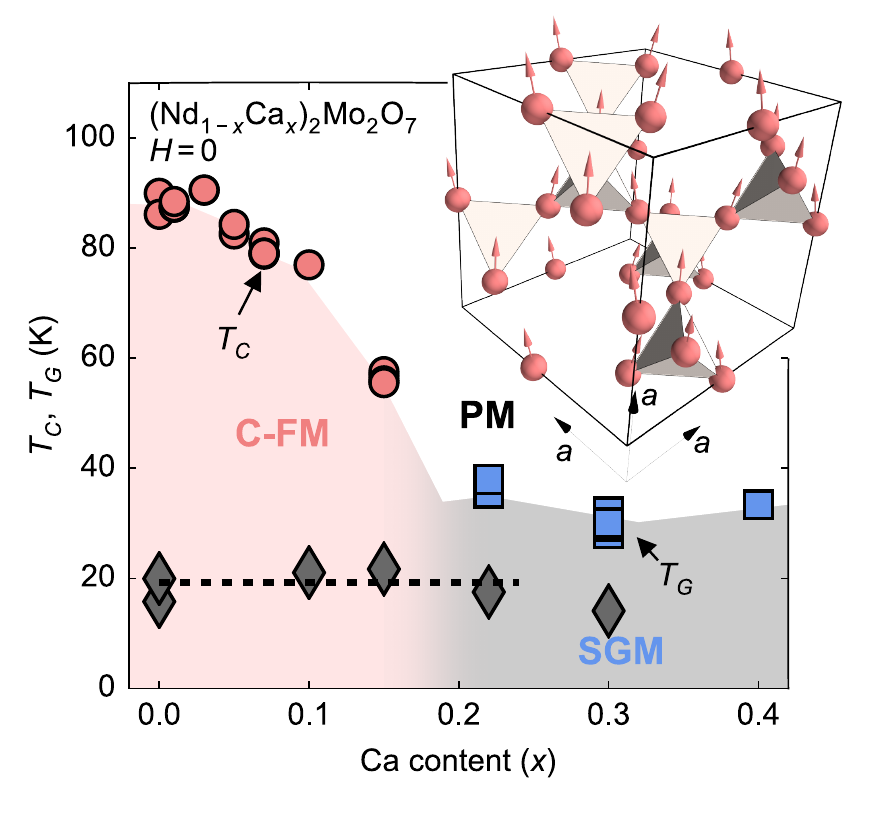}
  \caption{Phase diagram of (Nd$_{1-x}$Ca$_x$)$_2$Mo$_2$O$_7$ (NCMO) at zero magnetic field, with canted ferromagnetic (C-FM), spin-glass metal (SGM), and paramagnetic (PM) regime. Red and blue markers are for Curie-temperature $T_C$ and for the onset of glassy order ($T_G$), respectively. Black diamonds and a dashed line indicate the strong departure of field-cooled and high-field cooled magnetization curves (c.f. Fig. \ref{fig:mt}). Some data points at low $x$ adapted from Ref. \cite{Hirschberger2021}. Inset: Conducting molybdenum (red spheres) sublattice of NCMO, stressing the tetrahedron as a key structural motif. The canting angle of magnetic moments (red arrows) is exaggerated.}
\label{fig:pdiag}
\end{figure}

\begin{figure}[htb]
  \centering
  \includegraphics[width=0.8\linewidth]{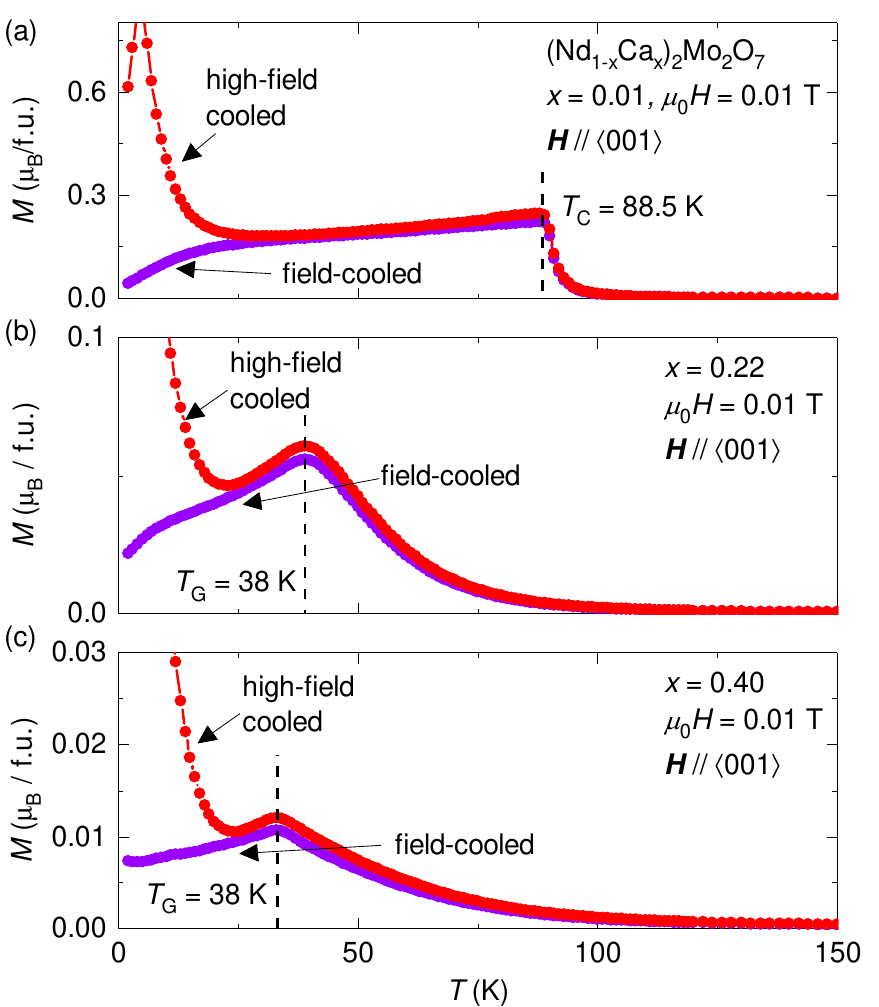}
  \caption{Low-field magnetization $M(T)$ in selected crystals of (Nd$_{1-x}$Ca$_x$)$_2$Mo$_2$O$_7$, for magnetic field along the $\left<001\right>$ magnetic easy-axis. Data was recorded for increasing temperature after cooling in $\mu_0 H = 1\,$T (red) and $\mu_0 H = 0.01\,$T (violet), respectively. Dashed vertical line indicates ordering transitions $T_C$, $T_G$ summarized in Fig. \ref{fig:pdiag}.}
\label{fig:mt}
\end{figure}

The phase diagram in Fig. \ref{fig:pdiag} summarizes the results of the $M-T$ measurements, including black diamonds at $T^*$ to mark the divergence of HFC and FC curves.

\subsection{Curie-Weiss analysis at high $T$: Mo-Mo interaction}
\label{ssec:cw_hight}
We further examine the magnetic susceptibility of NCMO at higher temperatures $T > \theta_\mathrm{Mo}\gg\theta_\mathrm{Nd}$ (dominant $d$-$d$ exchange), where the two sublattices can be decoupled at the mean-field level
\begin{equation}
\label{eq:si_chisum}
\chi_\mathrm{DC}  = \chi_\mathrm{Nd}+\chi_\mathrm{Mo} \approx \frac{C_\mathrm{Nd}}{T}+\frac{C_\mathrm{Mo}}{T-\theta_\mathrm{Mo}}
\end{equation}
As shown in section \ref{ssec:cw_lowt}, exchange interactions for the Nd$^{3+}$ moments are weak so that $\theta_\mathrm{Nd}/ T\ll 1$ is justified. The Curie-Weiss constants are positive and defined as 
\begin{equation}
\label{eq:cwlaw}
C_\alpha = \frac{\mu_0 \left(m_\alpha^\mathrm{CW}\right)^2\mu_B^2}{3k_B V_{m,\alpha}}
\end{equation}
where the fluctuating magnetic moments $m_\alpha^\mathrm{CW}$ ($\alpha =\,$Nd, Mo) are given in units of Bohr magneton and $V_{m,\alpha}$ is the volume assigned to each magnetic moment in the crystal. We use $V_{m,\alpha} = V_\mathrm{uc}/N$ with unit cell volume $V_\mathrm{uc}$, $N = N_\mathrm{fu}\cdot 2\cdot 2$, and $N_\mathrm{fu}=8$ is the number of formula units per unit cell. One factor of two accounts for two Nd (Mo) spins per formula unit, and the second $\times2$ represents the separation into two sublattices~\footnote{As an alternative to the separation into two sublattices, we may approximate the system as a single lattice where the moment is represented by an average value of Mo and Nd contributions. In this case, there are four magnetic moments per formula unit; thus motivating the additional factor $\times 2$ in the expression for the reference volume.}.

\begin{figure}[htb]
  \centering
  \includegraphics[width=0.8\linewidth]{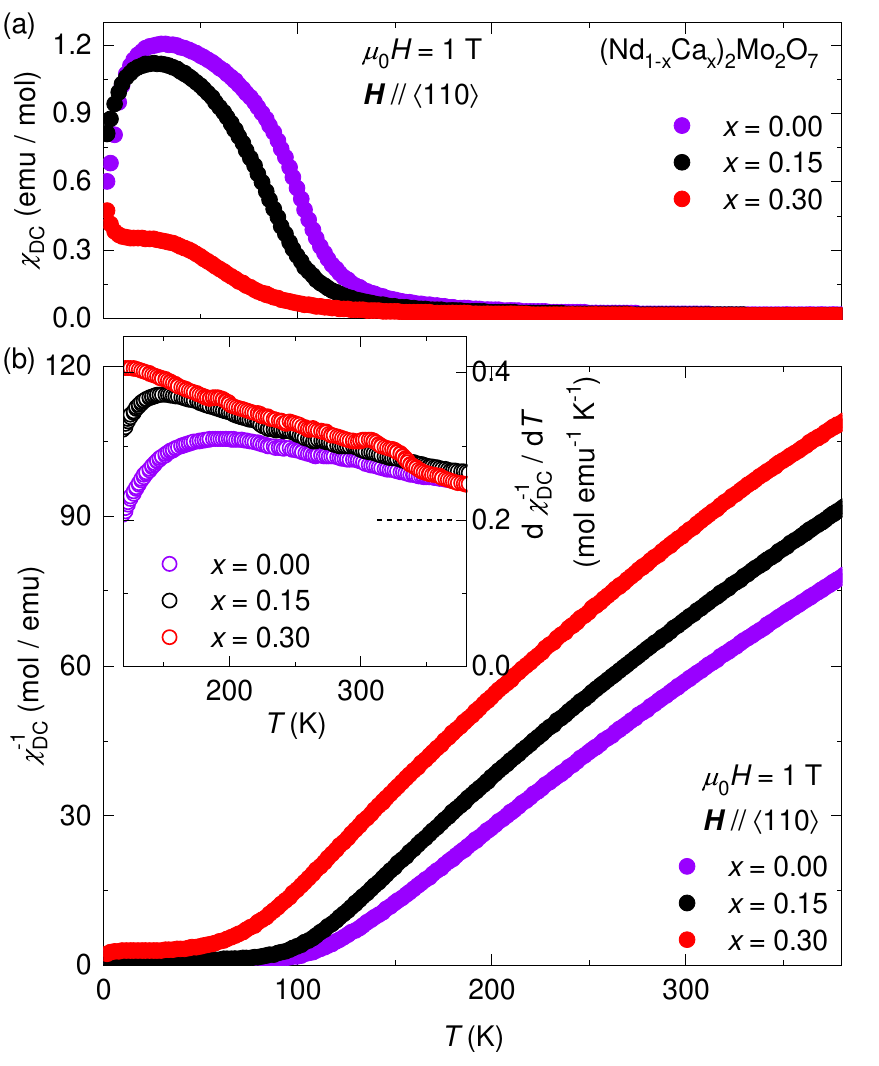}
  \caption{Curie-Weiss analysis of magnetic susceptibility $\chi_\mathrm{DC} = M/H$ in (Nd$_{1-x}$Ca$_x$)$_2$Mo$_2$O$_7$. (a) $\chi_\mathrm{DC}$ at elevated field of $\mu_0 H = 1\,$T for four selected crystals. (b) Inverse susceptibility retains finite curvature even above room temperature. The magnetic field was along the $\left<110\right>$-axis, but the behavior at these larger values of $H$ is nearly isotropic. Inset: Derivative of the inverse susceptibility with respect to temperature, with dashed line at right side of panel marking the $T\rightarrow \infty$ limit of $C_\mathrm{HT}^{-1}\approx d\chi_\mathrm{DC}^{-1}/dT$ predicted from the free-ion values for the magnetic moments. }
\label{fig:curie}
\end{figure}

\begin{figure}[htb]
  \centering
  \includegraphics[width=0.8\linewidth]{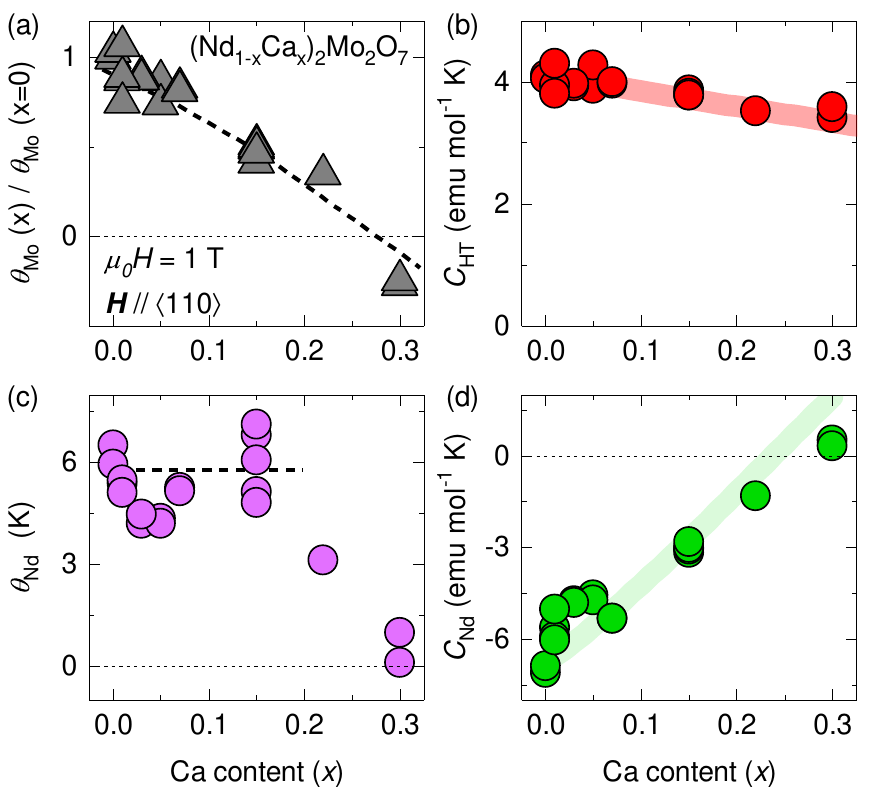}
  \caption{Curie-Weiss constants $C_\mathrm{HT}$, $C_\mathrm{Nd}$ and -temperatures $\theta_\mathrm{Mo}$, $\theta_\mathrm{Nd}$ extracted from susceptibility data as shown in Fig. \ref{fig:curie}. In (a,c), the dashed lines are guides to the eye. The shaded lines in (b,d) were calculated from Eq. (\ref{eq:c_ht}) and (\ref{eq:c_nd}), respectively.}
\label{fig:curie_analysis}
\end{figure}

To leading order in $\theta_\mathrm{Mo}/T$, Eq. (\ref{eq:si_chisum}) may be rewritten as
\begin{equation}
\label{eq:si_chisum_simple}
\chi_\mathrm{DC}  \approx \frac{C_\mathrm{HT}}{T-\theta_\mathrm{Mo}\left(C_\mathrm{Mo}/C_\mathrm{HT}\right)}
\end{equation}
with $C_\mathrm{HT}=C_\mathrm{Nd}+C_\mathrm{Mo}$. Notably, $C_\mathrm{Nd}$ depends on $T$ because the magnetic moment of Nd$^{3+}$ is suppressed upon cooling due to the crystal field effect. Figure \ref{fig:curie}(b) illustrates this point using the derivative $d\chi^{-1}/dT$, which retains finite curvature above room temperature. We conclude that a naive linear fit of $\chi_\mathrm{DC}^{-1}$ vs. temperature $T$ [Fig. \ref{fig:curie}(b)] provides only a rough estimate of the parameters $\theta_\mathrm{Mo}$ and $C_\mathrm{HT}$. Figure \ref{fig:curie_analysis}(a,b) summarizes the results of this fit procedure.

The red shaded line in Fig. \ref{fig:curie_analysis}(b) models the effect of Ca-doping on $C_\mathrm{HT}$ as follows. The parameters $m_\mathrm{Nd}^\mathrm{CW}$, $m_\mathrm{Mo}^\mathrm{CW}$, and $V_{m,\mathrm{Mo}}$ are kept unchanged, while enlarging the volume $V_{m,\mathrm{Nd}}(x) = V_{m,\mathrm{Nd}}(0)/(1-x)$ due to dilution of magnetic moments on the rare earth sublattice. Therefore,
\begin{equation}
\label{eq:c_ht}
C_\mathrm{HT}(x) = C_\mathrm{HT}(0)\times\frac{\left(m_\mathrm{Mo}^\mathrm{CW}\right)^2+\left(m_\mathrm{Nd}^\mathrm{CW}\right)^2\,(1-x)}{\left(m_\mathrm{Mo}^\mathrm{CW}\right)^2+\left(m_\mathrm{Nd}^\mathrm{CW}\right)^2}
\end{equation}
The agreement between this estimate and the observations is reasonable, and a corollary is that 
\begin{equation}
\frac{C_\mathrm{Mo}}{C_\mathrm{HT}}=\frac{\left(m_\mathrm{Mo}^\mathrm{CW}\right)^2}{\left(m_\mathrm{Mo}^\mathrm{CW}\right)^2+\left(m_\mathrm{Nd}^\mathrm{CW}\right)^2}
\end{equation}
at fixed temperature has rather weak dependence on $x$, while approaching $C_\mathrm{Mo}/C_\mathrm{HT}\rightarrow 0.40$ at $T\rightarrow \infty$ when the magnetic moments reach their respective free ion values, $m_\mathrm{Mo}^\mathrm{CW} = 2.83\,\mathrm{\mu_B/Mo}$ and $m_\mathrm{Nd}^\mathrm{CW} = 3.45\,\mathrm{\mu_B/Nd}$. Nevertheless, we found it preferable to consider only the relative change $\theta_\mathrm{Mo}(x)/\theta_\mathrm{Mo}(x=0)$ in Fig. \ref{fig:curie_analysis}(a). The $T\rightarrow \infty$ limit of $C_\mathrm{HT}^{-1}\sim d\chi_\mathrm{DC}^{-1}/dT$ is also marked by a black dashed line in Fig. \ref{fig:curie}(b), inset.

\subsection{Curie-Weiss analysis at low $T$: Mo-Nd interaction}
\label{ssec:cw_lowt}
In agreement with previous work on stoichiometric Nd$_2$Mo$_2$O$_7$, we observed that a tail appears in the $\chi_\mathrm{DC}(\mu_0H=1\,\mathrm{T})$ curves when Nd$^{3+}$ moments begin to order at $T<20\,$K. Figure \ref{fig:curie} shows that this contribution changes sign when the hole content is increased to $x = 0.30$. 

With the aim of extracting further information about the coupling between $4d$ conducting states and the rare earth moments, we fitted 
\begin{equation}
\chi_\mathrm{DC}=\chi_0+C_\mathrm{Nd}^{'}/\left(T-\theta_\mathrm{Nd}\right)
\end{equation}
at $T<20\,$K. Here, $\chi_0$ is the susceptibility of the Mo-$4d$ moments, known to be saturated in this temperature regime \cite{Mirebeau2007, Apetrei2007}. The resulting parameters are shown in Fig. \ref{fig:curie_analysis}(c,d). We found that $\theta_\mathrm{Nd}$, which measures effective Nd-Nd interactions $J_\mathrm{ff}$, is robust as a function of $x$ until the collapse of the canted ferromagnetic state at $x\approx 0.2$. 

The value of the Curie-Weiss constant $C_\mathrm{Nd}^{'}$ at $x=0$ can be used to obtain information about the $d$-$f$ coupling $J_\mathrm{df}$ between the two magnetic sublattices in Nd$_2$Mo$_2$O$_7$. To explain, let us emphasize that while $\chi_\mathrm{DC}$ of Fig. \ref{fig:curie} is calculated from $M/H$, $J_\mathrm{df}$ should be taken into account according to $M/\left(H+H_\mathrm{df}\right)$, where $H_\mathrm{df}$ is an exchange field at the Nd$^{3+}$ site in mean-field approximation:
\begin{equation}
\label{eq:curie_lt}
C_\mathrm{Nd}^{'}(0) = \frac{H+H_\mathrm{df}}{H}\,\frac{\mu_0 \left(m^\mathrm{CW}_\mathrm{Nd}\right)^2\mu_B^2}{3k_B V_{m,\mathrm{Nd}}}
\end{equation}
At low temperature, the magnetic moment of Nd$^{3+}$ is reduced by magnetic anisotropy and $m^\mathrm{CW}_\mathrm{Nd} = 2.35\,\mathrm{\mu_B/Nd}$~\cite{Taguchi2003}. Thus, from $C_\mathrm{Nd}^{'}(x=0)=-6.79\,\mathrm{emu\,K/mol}$ we deduce $\mu_0 H_\mathrm{df} = -10.9\,$T when $\mu_0H = 1\,$T. It is encouraging that this value is identical to, although of opposite sign than, the $H_\mathrm{df}$ felt by Gd$^{3+}$ ions in Gd$_2$Mo$_2$O$_7$ as detected by neutron scattering and the Moessbauer effect \cite{Mirebeau2007}.

In contrast to $\theta_\mathrm{Nd}$, $C_\mathrm{Nd}^{'}$ depends strongly on $x$ even when the hole content is low. Eq. \ref{eq:curie_lt} with adjusted reference volume $V_{m,\mathrm{Nd}}(x) = V_{m,\mathrm{Nd}}(0)/(1-x)$ yields
\begin{equation}
\label{eq:c_nd}
C_\mathrm{Nd}^{'}(x) =C_\mathrm{Nd}^{'}(0)\times(1-x)\times\frac{H+H_\mathrm{df}(x)}{H+H_\mathrm{df}(x=0)}
\end{equation}
Starting from the assumption that $H_\mathrm{df}\sim J_\mathrm{df}$ is roughly proportional to the dominant energy scale $J_\mathrm{dd}$, or $\theta_\mathrm{Mo}$, it follows $H_\mathrm{df}(x)/H_\mathrm{df}(0)\approx \theta_\mathrm{Mo}(x) / \theta_\mathrm{Mo}(0)$. The resulting green shaded line in Fig. \ref{fig:curie_analysis}(d) shows good agreement with the observations. Note that the measured $\theta_\mathrm{Nd}(x)$ cannot be described by merely assuming dilution of Nd$^{3+}$ spins on the A-sublattice of Nd$_2$Mo$_2$O$_7$ by Ca-doping; the variation of $H_\mathrm{df}$ is crucial. This confirms that $J_\mathrm{df}$ is closely correlated with the change of $J_\mathrm{dd}$ from predominantly ferromagnetic to predominantly antiferromagnetic. 

\section{Discussion}

Pyrochlore moybdates $R_2$Mo$_2$O$_7$ ($R=\,$ rare earth ion) are well known as a material platform for the study of competing, orbitally dependent double-exchange and magnetic superexchange interactions~\cite{Solovyev2003,Hanasaki2007,Motome2010a, Motome2010b}. Specifically in Gd$_2$Mo$_2$O$_7$, Nd$_2$Mo$_2$O$_7$, and Sm$_2$Mo$_2$O$_7$, a spin-glass metal (SGM) state was reported to appear when enhancing superexchange interactions~\cite{Biermann2005,Apetrei2006, Mirebeau2007, Apetrei2007,Iguchi2009} by means of hydrostatic pressure~\cite{Iguchi2009}. The present SGM at $x>0.22$ resembles the pressure-induced phase both in terms of the cusp-like magnetic susceptibility $\chi(T)$ and a slightly semiconducting resistivity curve (not shown) with absolute value close to the Mott-Ioffe-Regel limit of hopping conduction. Figure \ref{fig:curie_analysis}(a) further demonstrates that the phase transition between C-FM and SGM is driven by a transition of $J_\mathrm{dd}$ from predominantly ferromagnetic to mostly antiferromagnetic. Regarding Mo-$R$ correlations, a sign change $J_\mathrm{df}$ upon entering the SGM state is consistent with neutron scattering studies on Gd$_2$Mo$_2$O$_7$ and (Tb$_{1-x}$La$_x$)$_2$Mo$_2$O$_7$ under pressure~\cite{Apetrei2006, Mirebeau2007,Apetrei2007}. Finally, effective ferromagnetic $J_\mathrm{ff}<0$ as deduced from $\theta_\mathrm{Nd}>0$ in Fig. \ref{fig:curie_analysis}(c), has universally been reported in $R_2$Mo$_2$O$_7$ for various $R$ using small-angle neutron scattering \cite{Mirebeau2007,Apetrei2007}. A neutron scattering experiment on Tb$_2$Mo$_2$O$_7$ under hydrostatic pressure further evidenced a reduction of $J_\mathrm{ff}$ by more than one third, but the complete collapse of $J_\mathrm{ff}$ at $x = 0.30$ as found here warrants further investigation.

The focus of the present study however is the canted ferromagnetic (C-FM) order at lower $x$, and the evolution of magnetic interactions within this phase. In Ref. \cite{Hirschberger2021}, electronic transport data in this regime was analyzed under the assumption that the canting angle $\alpha_\mathrm{Mo}$ of Molybdenum spins in the low-field region does not change dramatically up to $x = 0.10$. We consider this problem in the context of the magnetization and magnetic susceptibility data presented above.

\begin{figure}[htb]
  \centering
  \includegraphics[width=0.8\linewidth]{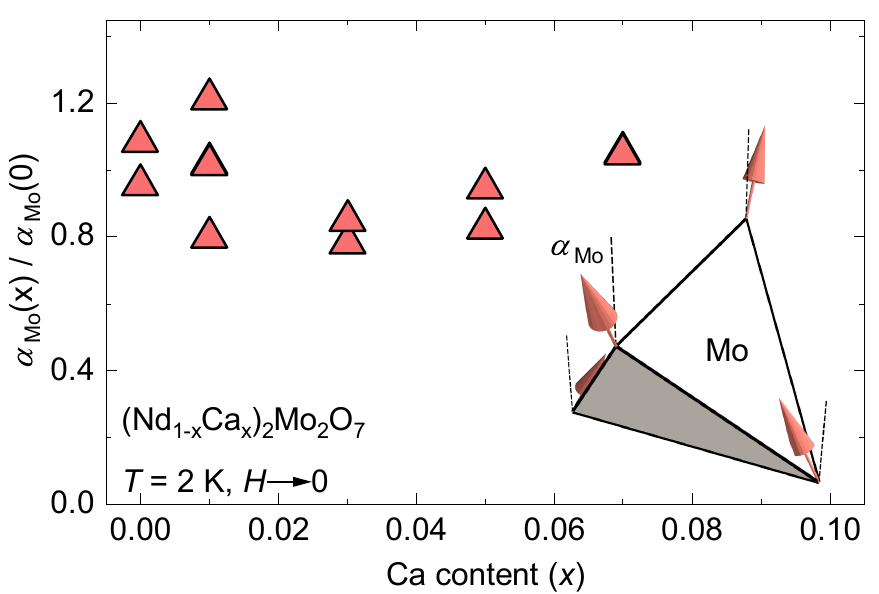}
  \caption{Canting angle $\alpha_\mathrm{Mo}$ of Molybdenum spins away from the $\left<001\right>$ direction. $\alpha_\mathrm{Mo}$ was calculated from the exchange integrals $J_\mathrm{dd}$, $J_\mathrm{df}$, which in turn originate from $\theta_\mathrm{Mo}$ and $C_\mathrm{Nd}$, respectively. Inset: Molybdenum magnetic moments (red arrows) on an individual tetrahedron, with dashed lines marking $\left<001\right>$.}
\label{fig:alphamo}
\end{figure}

Figure \ref{fig:curie_analysis} demonstrates the evolution of exchange constants $J_\mathrm{dd}$ and $J_\mathrm{df}$ as mirrored by the observables $\theta_\mathrm{Mo}$ and $C_\mathrm{Nd}^{'}$. $J_\mathrm{dd}\sim \theta_\mathrm{Mo}$ and $J_\mathrm{df}\sim H_\mathrm{df}$ decrease in unison with hole content $x$, their ratio being essentially constant at $x\le 0.15$. Canting in the Mo sublattice emerges from the competition of these two energy scales: At $x = 0$, we estimate $J_{dd} = (3/2z)k_BT_C/(S(S+1))\approx 11\,$K with $S=1$ for Mo and $z=6$ the coordination number, i.e. number of nearest neighbor Molybdenum sites in the pyrochlore lattice. Likewise, $J_\mathrm{df} = 1.5\,$K can be estimated from $H_\mathrm{df}$ reported above. Although the absolute value of the canting angle cannot be determined with a high precision using these considerations, $\alpha_\mathrm{Mo}=5-10^\circ$ was previously determined from neutron scattering experiments~\cite{Taguchi2001,Yasui2003}. We can use $\alpha_\mathrm{Mo}\sim J_\mathrm{df}/J_\mathrm{dd}$ to determine the $x$-dependence of $\alpha_\mathrm{Mo}$ in Figure \ref{fig:alphamo}.

Figure \ref{fig:alphamo} hence demonstrates the main result of our study, namely the robustness of $\alpha_\mathrm{Mo}$ within the C-FM phase. This conclusion is also supported by the unchanged nature of $\rho_{xx}(T = 2\,$K$, H =0)$ reported in Ref. \cite{Hirschberger2021}. Hopping is suppressed in $R_2$Mo$_2$O$_7$ when tilting $\alpha_\mathrm{Mo}>0$, so that $\rho_{xx}$ is expected to be a highly sensitive probe of the relative angle between neighboring Mo spins \cite{Moritomo2001,Solovyev2003,Hanasaki2007,Iguchi2009,Motome2010}. 

\section{Conclusion and Outlook}
Above, we have studied high quality single crystals of the (Nd$_{1-x}$Ca$_x$)$_2$Mo$_2$O$_7$ (NCMO) series up to $x=0.40$ using x-ray and magnetization techniques. Our analysis is capable of separating contributions to the magnetic susceptibility from two sublattices, and we extract a hierarchy of effective spin-spin interactions in this solid solution: $J_\mathrm{dd}$ (Mo-Mo), $J_\mathrm{df}$ (Mo-Nd), and $J_\mathrm{ff}$ (Nd-Nd). Comparing the extracted values to magnetization and previous neutron scattering data, it is shown that the local-ion anisotropy on the Mo site with easy $\left<001\right>$ direction is weak; a small canting angle $\alpha_\mathrm{Mo}$ is imparted on the 4$d$ sublattice due to a combination of $J_\mathrm{df}$ and the noncoplanar (two in, two out) nature of the Nd$^{3+}$ sublattice. Most importantly, the lattice-averaged $\alpha_\mathrm{Mo}$ is shown to be robust when introducing Ca$^{2+}$ on the Nd$^{3+}$ site. Our findings are noteworthy when considering that advanced experimental techniques such as elastic neutron scattering should struggle to resolve small changes in $\alpha_\mathrm{Mo}\approx 5-10^\circ$, especially due to the presence of the Nd-sublattice with significantly much larger moment size and stronger noncollinearity~\cite{Yasui2001,Yasui2003,Yasui2006}.

The role of spin-orbit coupling (SOC) in pyrochlore molybdates $R_2$Mo$_2$O$_7$ remains a mystery warranting further investigation. Given the weak magnetic anisotropy of Molybdenum in these spin $S=1$ compounds, the orbital moment is believed to play a minor role~\cite{Solovyev2003}. Nevertheless, new theories have suggested significant SOC in at least some of the $R_2$Mo$_2$O$_7$, particularly on the insulating side of the Mott transition ($R=$ Ho, Y, Dy, Tb)~\cite{Silverstein2014,Thygesen2017,Li2018,Smerald2019}. Meanwhile, recent measurements of the Hall effect in the field-aligned ferromagnetic regime of NCMO revealed a large, SOC-driven signal with strong dependence on the position of the chemical potential $\varepsilon_F$~\cite{Hirschberger2021}.

The present study thus demonstrates the robust canting angle $\alpha_\mathrm{Mo}$ for the conducting Molybdenum states at moderate Ca-content in (Nd$_{1-x}$Ca$_x$)$_2$Mo$_2$O$_7$ and establishes this pyrochlore oxide as a promising target for studies of Berry-curvature physics at the interface of spin-noncoplanarity and SOC~\cite{Zhang2020, Lux2020, Kipp2020}.

\textit{Acknowledgments.} We acknowledge discussions with Y. Taguchi and K. Ueda. This work was partially supported by JST CREST Grant Number JPMJCR1874 (Japan). 

\bibliography{NCMO_Magnetism}

\begin{figure}[htb]
  \centering
  \includegraphics[clip, trim=0.35cm 0.36cm 0.72cm 0.9cm, width=0.7\linewidth]{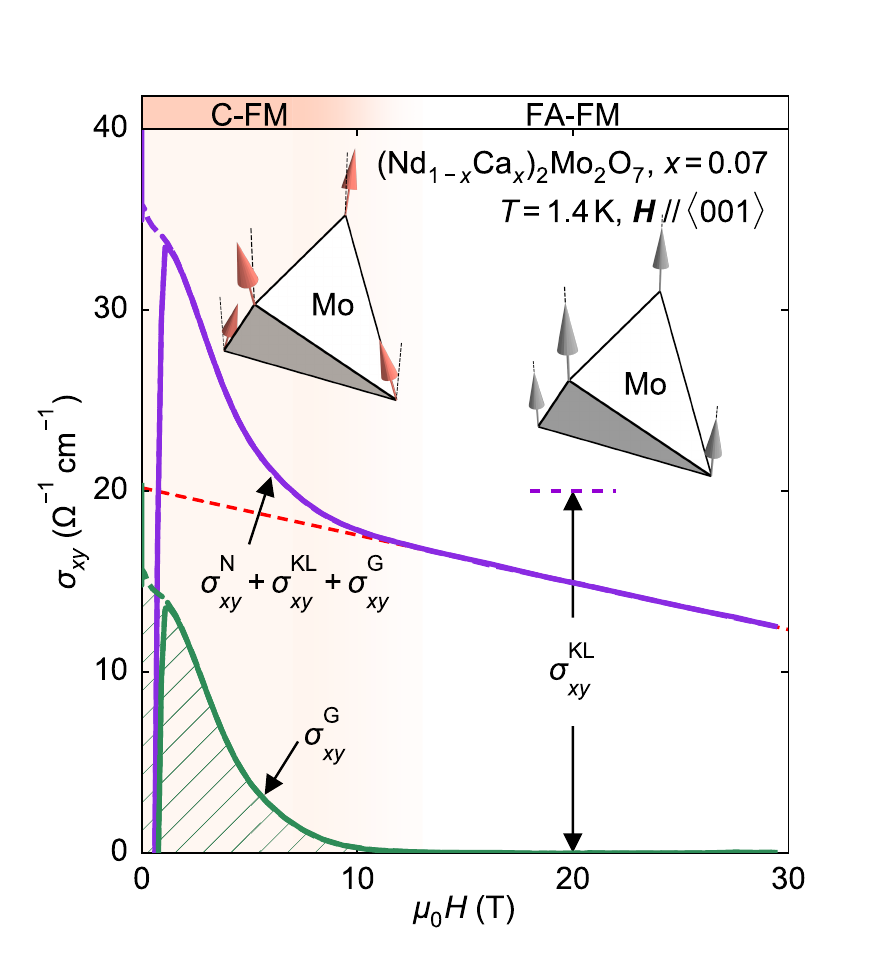}
  \caption{Unconventional Hall effect from momentum-space Berry curvature in (Nd$_{0.93}$Ca$_{0.07}$)$_2$Mo$_2$O$_7$. The total Hall conductivity (violet) is separated into three components: normal Hall effect $\sigma_{xy}^\mathrm{N}$ (slope of red dashed line), spin-orbit driven anomalous Hall effect $\sigma_{xy}^\mathrm{KL}$, and the geometrical Hall conductivity $\sigma_{xy}^\mathrm{G}$ (green line, hatched area). The latter is observed only in the low-field noncoplanar ground state, termed canted ferromagnetic phase (C-FM, orange shading and left inset). $\sigma_{xy}^\mathrm{KL}$ is present both in C-FM and the field-aligned ferromagnetic state (FA-FM, right inset). Insets show canted (left) and collinear (right) configurations of Molybdenum moments on an individual tetrahedron of the pyrochlore structure. Hall effect data adapted from Ref. \cite{Hirschberger2021}.}
\label{fig:hall}
\end{figure}

\end{document}